*Original Article*

# Effect on New Loan Repayment Fine Clause on Bank Jaya Artha's Customer Satisfaction and Recommendation

[1]**Mustaqim Adamrah**, [2]**Yos Sunitiyoso**

[1,2]*Decision Making and Strategic Negotiation Department, Master of Business Administration, School of Business and Management Institut Teknologi Bandung, Jakarta, Indonesia.*

***Abstract:*** *The growing population of older people in Indonesia—the world's fourth-most populous country—makes a larger cake for the pension business, including in the banking sector. PT Bank Jaya Artha is one of the Indonesian banks that provide products and services for people who are set to enter retirement age. In the wake of tight competition in the pension business market, Bank Jaya Artha has since 2019 imposed a fine of three times of installments in addition to 5% of outstanding debt on customers planning to repay their debt in a mission to prevent them from leaving for competitors. While the clause is not included in loan agreements signed before the implementation, it applies to past loan agreements as well. This, in turn, has led to customer complaints. The research is meant to find out how the implementation of the unconsented clause has affected customer satisfaction and willingness to recommend the bank and what the bank should do to become more customer-centric, according to customers. Using a design thinking framework, the research collects quantitative and qualitative data from the bank's pension customers through questionnaires and forum group discussions. Statistical analysis is utilized on quantitative data from questionnaires, and content analysis is utilized on qualitative data from questionnaires. A narrative analysis is also used to explain qualitative data from forum group discussions. The result shows that there are problems in the way the bank communicates information to customers, particularly information about the loan repayment fine. Lack of transparency, a reactive approach instead of a proactive one, the obscurity of the information, and the time the information is delivered have affected customers' satisfaction toward the bank.*

## I. INTRODUCTION

Indonesia, the largest economy in Southeast Asia by gross domestic product, offers a great potential market to businesses due to its huge population. In the financial sector, banks—national or foreign alike, regional or national alike, conventional or sharia alike—are racing to grab a slice of a huge cake by providing a range of products and services to different customer segments that may suit their customers' financial needs.

PT Bank Jaya Artha is competing with more than 10 other national and regional banks, including state-owned ones, in targeting the growing population of older people.

**Table 1: Population of Indonesians in Retirement Age Ranges**

| Age Group | 2020 | | 2021 | | 2022 | |
|---|---|---|---|---|---|---|
| | '000 | % | '000 | % | '000 | % |
| 55-59 | 13,120.90 | 4.9% | 13,531.70 | 5.0% | 13,961.20 | 5.1% |
| 60-64 | 10,209.50 | 3.8% | 10,617.80 | 3.9% | 11,061.50 | 4.0% |
| 65-59 | 7,454.00 | 2.8% | 7,828.50 | 2.9% | 8,199.00 | 3.0% |
| 70-74 | 4,553.90 | 1.7% | 4,892.40 | 1.8% | 5,269.40 | 1.9% |
| 75+ | 4,624.50 | 1.7% | 4,860.10 | 1.8% | 5,130.60 | 1.9% |
| **Total in Indonesia** | 270,203.90 | 14.8% | 272,682.50 | 15.3% | 275,773.80 | 15.8% |

*Note: The result of Interim Population Projection 2020-2023 (mid-year/June)*
*Source: The National Statistics Agency (BPS) [3]*

In the wake of tight competition in the pension business market, Bank Jaya Artha has implemented an unconsented clause that requires customers to pay a fine of 5% of outstanding debt and another fine of three times of installment should they plan to repay their loan before maturity.

The clause, which was introduced in December 2019, applies to loan agreements signed before and after the date of the implementation. This has caused different customers to file similar complaints regarding the clause through the bank's contact center, mass media, social media, a regional legislative council where they reside, or the Financial Services Authority. In some cases, customers filed their complaints against the bank with a district court. Bank Jaya Artha's pension business unit already reported 828 complaints as of July 2023, exceeding the full-year figure of 821 recorded in 2022.

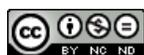

Table 2: Complaints at Bank Jaya Artha's Pension Business Unit

|  | 2018 | 2019 | 2020 | 2021 | 2022 | 2023 (as of July) |
|---|---|---|---|---|---|---|
| Complaints on Loan Repayment Fine | 2 | 3 | 32 | 38 | 17 | 9 |
| Complaints on the Loan Repayment Process | 36 | 66 | 50 | 119 | 62 | 64 |
| Total Complaints at Pensions Business | 1,233 | 1,449 | 1,907 | 1,093 | 821 | 828 |

*Source: Bank Jaya Artha's Internal Data*

The research is to look for answers as to how the implementation of the unconsented clause has affected customer satisfaction and willingness to recommend the bank and what the bank should do to become more customer-centric.

The scope of the research covers only Bank Jaya Artha's pension business unit and the unconsented clause. Primary data comes from the bank's data, policy, questionnaires, and forum group discussions. The research is limited to pension customers in Jakarta, Tangerang, Bekasi, Bogor, Semarang, and Surabaya through questionnaires and to selected pension customers in Jakarta through forum group discussions.

## II. LITERATURE REVIEW

Customer behavior theories can be used to explore and explain business issues in Bank Jaya Artha's pension business unit to understand customer reactions to the loan agreement clause. Customers behave in a way that responds to the provider of products or services they buy and expect something from the provider. Any action carried out by the provider could affect customers' emotions in customer complaints and, eventually, their decision-making.

Utpal M. Dholakia is a George R. Brown Professor of Marketing at Rice University's Jesse H. Jones Graduate School of Business said in his article published by Harvard Business Review on June 29, 2021 (https://hbr.org/2021/06/if-youre-going-to-raise-prices-tell-customers-why) that a company's communication in a price increase, when performed poorly, can lead to undesirable outcomes like customer complaints, social media outrage, and even worse, having to walk back the price increase, or losing customers altogether.

In another article published by Harvard Business Review's December 2007 issue (https://hbr.org/2007/12/the-customers-revenge), Dan Ariely—the writer of the best-selling book Predictably Irrational: The Hidden Forces That Shape Our Decisions—said customers perceive interactional fairness when they believe they have been treated with respect and empathy and they have been given reasonable explanations for actions taken.

He also said: "Most unhappy customers do not bother to publicize their experiences ... Instead, they stop buying and spread bad news on their social networks. These small revenges represent a great loss of lifetime value that is invisible to the firm but has substantial implications for the bottom line."

Stakeholder theory can also be used to analyze the interests and concerns of different stakeholders. Combined ideas of stakeholder theory (Freeman, 1984; Freeman et al., 2010) create a new form that says: "Business is a set of value-creating relationships among groups that have a legitimate interest in the activities and outcomes of the firm and upon whom the firm depends to achieve its objectives. It is about how customers, suppliers, employees, financiers (stockholders, bondholders, banks, etc.), communities, and management work cooperatively to create value. Understanding a business means understanding how these relationships work."

The research uses design thinking as a conceptual framework to understand customers better. In the first step of design thinking, or the empathize phase, the researcher learnt to understand who the target group of this research was, their background, their routine, their preference of way of communication, and their behavior, based on observations—including from complaints pension customers filed with the bank—questionnaires at six cities, and answers at forum group discussions.

In the define step, the researcher identified customers' needs and problems from answers they provided through questionnaires and questions at forum group discussions and then asked for customers' feedback to find solutions in the ideate phase. The researcher limited the process to the ideate phase only due to the limited time of research.

The researcher collected quantitative and qualitative data by asking 162 walk-in customers at branches in Jakarta, Tangerang, Bekasi, Bogor, Semarang, and Surabaya through questionnaires developed from customer satisfaction surveys. The researcher also collected qualitative data from 11 selected respondents in Jakarta through two sessions of forum group discussions.

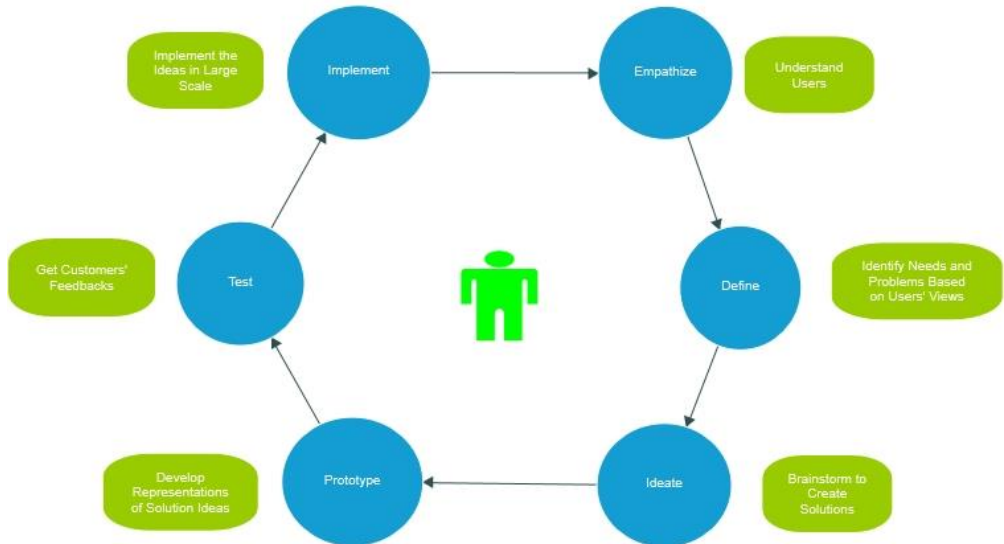

**Figure 1: Conceptual Framework of Design Thinking**

**Table 3: Customers' Data Collected in Questionnaires**

| Basic Information | Main Questions |
|---|---|
| Gender | General information delivery satisfaction |
| Age | Loan ownership |
| City of origin | Knowledge about loan repayment fine |
| Length of Engagement | Loan repayment fine information delivery satisfaction |
| | Overall satisfaction |
| | Input on communications |
| | Input on services |
| | Recommendation likeliness |

**Table 4: Information Collected in Forum Group Discussions**

| Basic Information | Main Questions |
|---|---|
| Gender | Attitude and behavior upon retirement |
| Age | Retirement finance conditions |
| Length of engagement | Interaction with Bank Jaya Artha |
| Occupational background | Communications from Bank Jaya Artha |
| Daily activities | Experience with Bank Jaya Artha |
| | Overall satisfaction |
| | Recommendation likeliness |

Quantitative data from questionnaires was processed using the Net Promoter Score (Reichheld, 2003; Markey and Reichheld, 2011). Qualitative data from questionnaires was made into groups to see the distribution of each group. The researcher used statistical analysis on quantitative data from questionnaires and content analysis on qualitative data from questionnaires. The researcher also used narrative analysis to explain qualitative data from forum group discussions.

### III. RESULTS AND DISCUSSION

The researcher distributed questionnaires to 162 walk-in pension customers at Bank Jaya Artha's branches in Jakarta, Tangerang, Bogor, Bekasi, Semarang, and Surabaya. A summary of statistics from questionnaires is presented in the table below.

**Table 5: Summary of Statistics from Questionnaires**

| Description | N | Mean/Proportion | Data Distribution | Min | Max |
|---|---|---|---|---|---|
| Gender (Female = 1) | 162 | 0.605 | 0.490 | 0 | 1 |
| Age | | | | | |
| Up to 60 Years Old | 162 | 0.105 | | | |
| 61-65 Years Old | 162 | 0.185 | | | |
| 66-70 Years Old | 162 | 0.265 | | | |
| > 70 Years Old | 162 | 0.444 | | | |
| **City** | | | | | |
| Bekasi | 162 | 0.080 | | | |
| Bogor | 162 | 0.185 | | | |
| Jakarta | 162 | 0.210 | | | |
| Semarang | 162 | 0.265 | | | |
| Surabaya | 162 | 0.216 | | | |
| Tangerang | 162 | 0.043 | | | |
| **Length of Engagement** | | | | | |
| Up to 5 Years | 162 | 0.216 | | | |
| 6-10 Years | 162 | 0.167 | | | |
| 11-15 Years | 162 | 0.296 | | | |
| > 15 Years | 162 | 0.321 | | | |
| General Information Delivery Satisfaction Score | 162 | 8.204 | 1.517 | 1 | 10 |
| Loan Ownership (Yes = 1) | 162 | 0.728 | 0.446 | 0 | 1 |
| Knowledge about Fine on Accelerated Loan Repayment (Yes = 1) | 118 | 0.297 | 0.459 | 0 | 1 |
| Fine Information Delivery Satisfaction Score | 118 | 4.992 | 2.281 | 1 | 10 |
| Overall Satisfaction Score | 162 | 8.556 | 1.290 | 5 | 10 |
| Recommendation Score | 162 | 6.784 | 2.642 | 1 | 10 |

*Source: Primary data*

Based on questionnaire results, female respondents dominated with 98 out of 162, or around 60.5% of the total samples. The highest number of respondents were in Semarang with 26.5%, followed by 21.6% respondents in Surabaya, 21.0% respondents in Jakarta, 18.5% respondents in Bogor, 8% respondents in Bekasi, and 4.3% respondents in Tangerang.

In terms of age, respondents over 70 years old were the majority, with 44.4%, followed by 26.5% of respondents aged 66-70 years old, 18.5% of respondents aged 61-65 years old, and 10.5% of respondents aged up to 60 years old. Around 32.1% of respondents have been engaged with the bank for over 15 years, while 29.6% of respondents have been with the bank for 11-15 years, 21.6% for up to five years, and 16.7% for 6-10 years.

In this research, questionnaire respondents were asked to give scores from 1 to 10 for a number of variables, such as how the bank communicated information to customers in general, how the bank communicated information about loan repayment fines, how satisfied customers were in overall aspects, and how likely customers would recommend the bank to someone else. For every question of those variables, the researcher also asked questionnaire respondents the reason behind the scoring.

In Net Promoter Score, 1-6 are considered detractors, 7-8 are considered passives, and 9-10 are considered promoters (Reichheld, 2003). As an end result, Net Promoter Score can range between -100% and 100%, and general approval is that anything above 0 is considered good, and the higher the number, the better. Voice-of-customer platform SurveySensum Head of Content Marketing Manisha khandelwal said a net promoter score of above 50% is great, and above 70% is excellent (surveysensum.com, 2023).

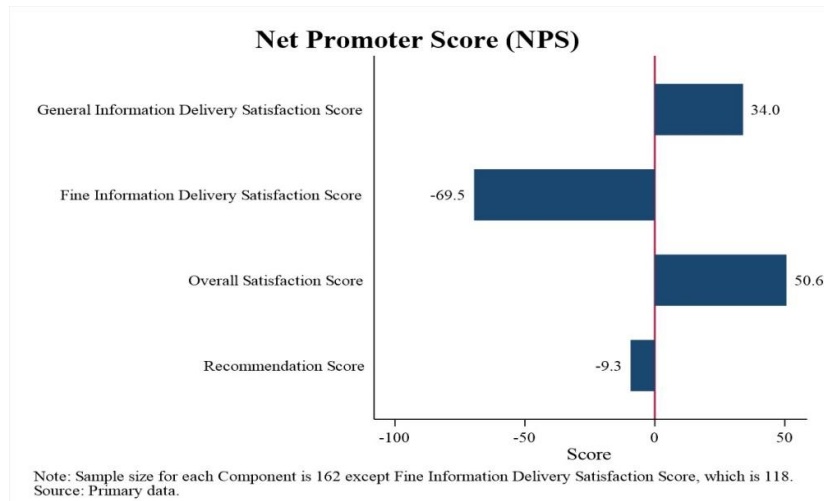

**Figure 2: Net Promoter Scores for Questionnaire Quantitative Variables**

**Table 6: Net Promoter Score Distribution**

| Score (1 = Lowest; 10 = Highest) | General Information Delivery Satisfaction | Fine Information Delivery Satisfaction | Overall Satisfaction | Recommendation |
|---|---|---|---|---|
| 1 | 1.23 | 8.47 | 0.00 | 8.64 |
| 2 | 0.00 | 5.93 | 0.00 | 1.85 |
| 3 | 0.00 | 14.41 | 0.00 | 0.62 |
| 4 | 0.00 | 8.47 | 0.00 | 5.56 |
| 5 | 3.09 | 22.03 | 2.47 | 16.67 |
| 6 | 4.32 | 17.80 | 3.70 | 2.47 |
| 7 | 17.90 | 9.32 | 15.43 | 12.35 |
| 8 | 30.86 | 5.93 | 21.60 | 25.31 |
| 9 | 20.37 | 4.24 | 27.78 | 10.49 |
| 10 | 22.22 | 3.39 | 29.01 | 16.05 |
| NPS (%) | 33.95 | -69.49 | 50.62 | -9.26 |
| N | 162 | 118 | 162 | 162 |

*Source: Primary Data*

Respondents gave an average score of 8.2 for how they perceived the bank in communicating information to customers in general. And when their scores are calculated in Net Promoter Score, it results in 34.0%, which is considered good (surveysensum.com, 2023).

The majority of the respondents, or 57.4% of the total, agreed that clarity of information was the main reason for their satisfaction with how the bank communicated its information in general to customers. Meanwhile, 25.3% of respondents said they were satisfied because the information given was sufficient, and 13.0% had their information needs fulfilled by the bank when they asked questions.

Respondents also gave negative sentiments, saying there was a lack of an information piece that the bank failed to deliver (10.5%), or the information they received from the bank was obscure (3.1%), or the bank only communicated important information to customers when asked (1.2%).

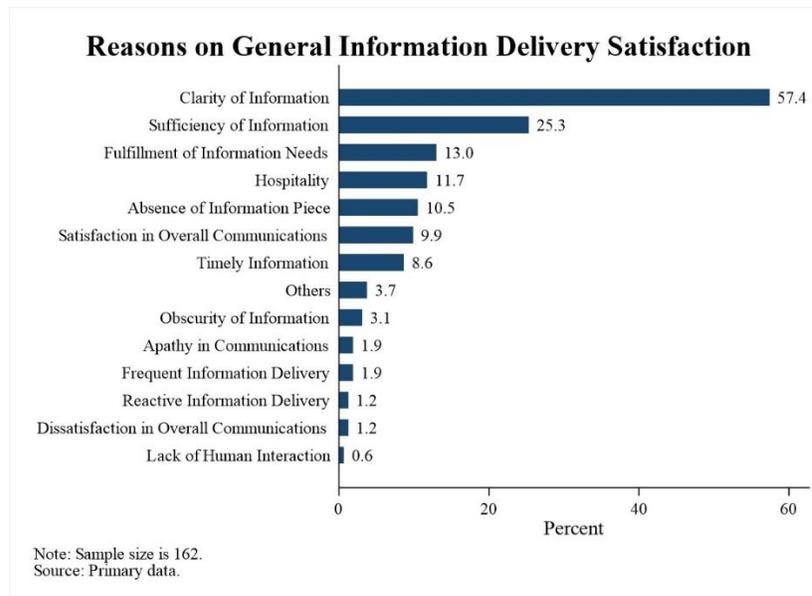

**Figure 3: Reasons for General Information Delivery Satisfaction**

Of 162 respondents, only 118 had loans with the bank. Therefore, only the 118 respondents were asked about their knowledge of loan repayment fines, whether or not they received the information about loan repayment fines, and how they gave scores from 1 to 10 for the delivery of loan repayment fines information. Of the 118 respondents who had loans with the bank, more than 70% said they did not know about the loan repayment fine clause. The 118 respondents gave an average score of 4.99 when asked about their opinion on how the bank informed the loan repayment fine to customers. The Net Promoter Score for loan repayment fine information delivery satisfaction stood at -69.5%, which means the bank needs improvement.

When asked for the reasons behind the fine information delivery satisfaction score, 48.1% of the 118 respondents said they had never received the information from the bank, while 14.8% of the loan owner respondents said they only learned the information when they asked the bank's staff upon their plan to repay their loans or when they heard from another customer who complained about the clause. Only 7.4% of the loan owner respondents said they received clear information about the loan repayment fine, only another 3.1% said the bank provided sufficient information, and only 2.5% said the bank provided the information in a timely manner before the signing of their loan agreements.

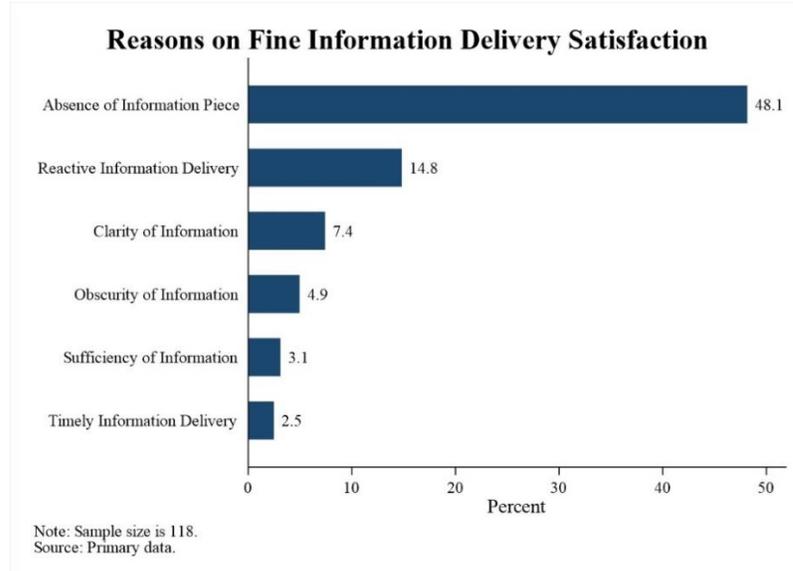

**Figure 4: Reasons for Fine Information Delivery Satisfaction**

When asked about overall satisfaction, 162 respondents gave an average score of 8.56, and the Net Promoter Score for this variable was 50.6%, which is considered great (surveysensum.com, 2023).

Around 35.8% of 162 respondents said they were satisfied overall because the bank's staff provided excellent services and helped customers get what they needed, 23.5% applauded the bank's speedy process of delivering services, and 20.4% said they felt welcomed by the bank's staff's hospitality. Respondents also provided negative feedback, such as 4.9% of respondents who were unsatisfied with the bank's loan repayment fine, and 4.3% who said there certain information lacked from the bank.

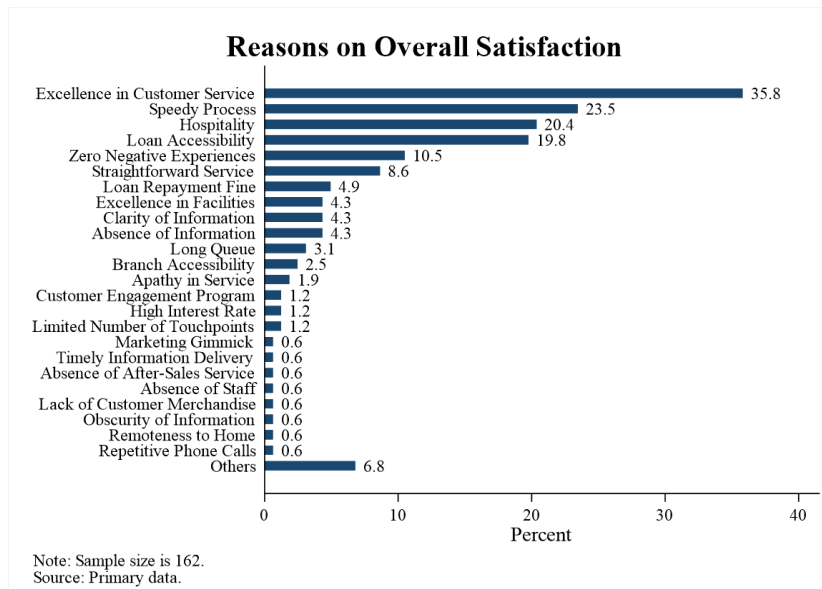

**Figure 5: Reasons for Overall Satisfaction**

When it comes to how likely customers would recommend the bank, the respondents scored an average of 6.78 and gave a Net Promoter Score of -9.3%. Only 26.5% of respondents would likely actively recommend the bank, according to Net Promoter Score for recommendations. Around 29.0% of respondents said the bank's excellent customer service was the main reason to recommend the bank, 21% said they might recommend the bank because it was easy to get loans from the bank, and 9.3% because the bank provided speedy processes. Around 16.7% of respondents said it was every individual's right to choose their own pension bank, while 14.2% of respondents said they had a limited social circle, which made them unlikely to recommend the bank.

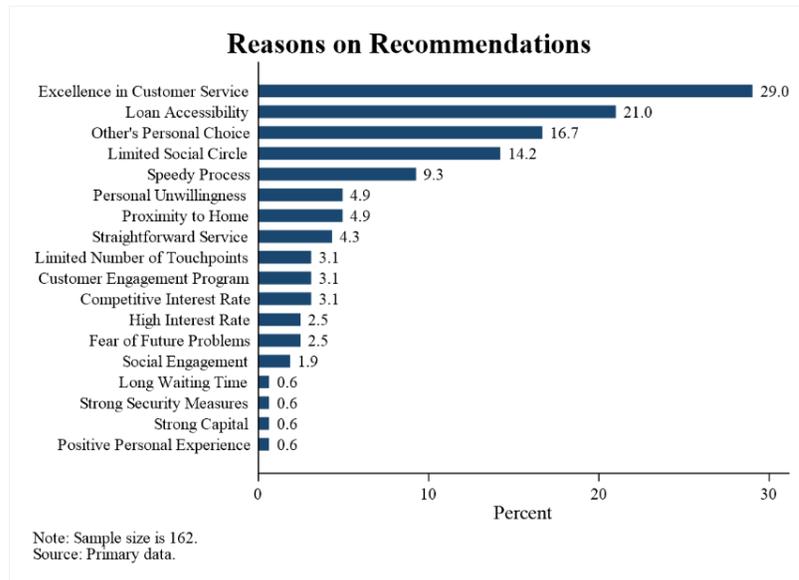

**Figure 6: Reasons for Recommendations**

The researcher also asked for questionnaire respondents' inputs for improvement in the bank's communications. While 113 out of 162 respondents provided no inputs, it is worth noting that 8.6% of respondents expected a more proactive approach from the bank in delivering information that could affect them. Around 8.6% of respondents also expected more empathetic communications from the bank's staff to pension customers, which requires the bank's staff to better understand customers' conditions and needs. As many as 6.8% of respondents said the bank needs to provide more details in delivering information, while 6.2% of respondents said the bank needs to be more transparent in delivering information.

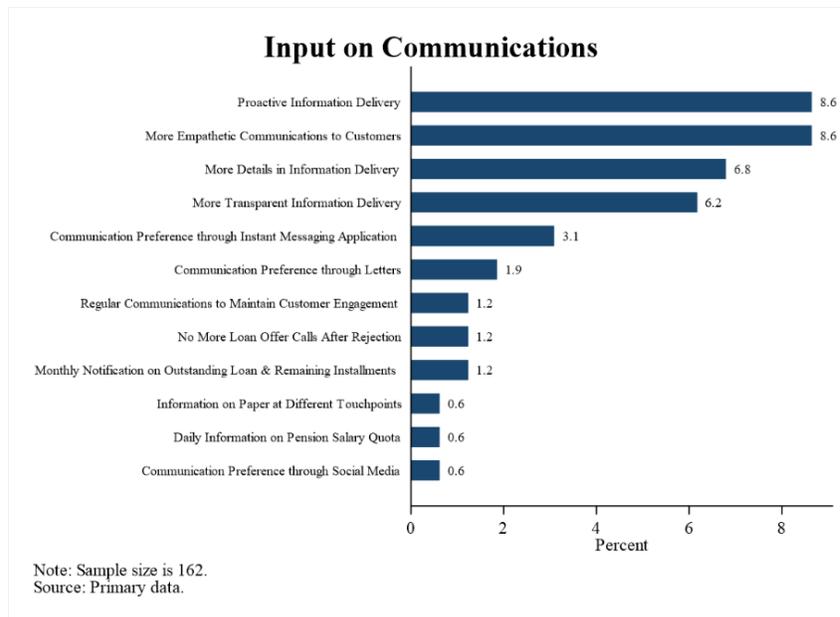

**Figure 7: Input on Communications**

On the service front, 106 of 162 respondents provided no input for improvement. As many as 12.3% of respondents expected the bank to cut waiting time, some of whom said that although the queue was short, waiting time could take longer than expected.

As many as 11.7% of respondents said they expected shorter queues, especially during the first few days of every month when they withdrew pension salaries. Around 8.6% of respondents said they expected more staff members at counters at the branch offices to help reduce waiting time and queues.

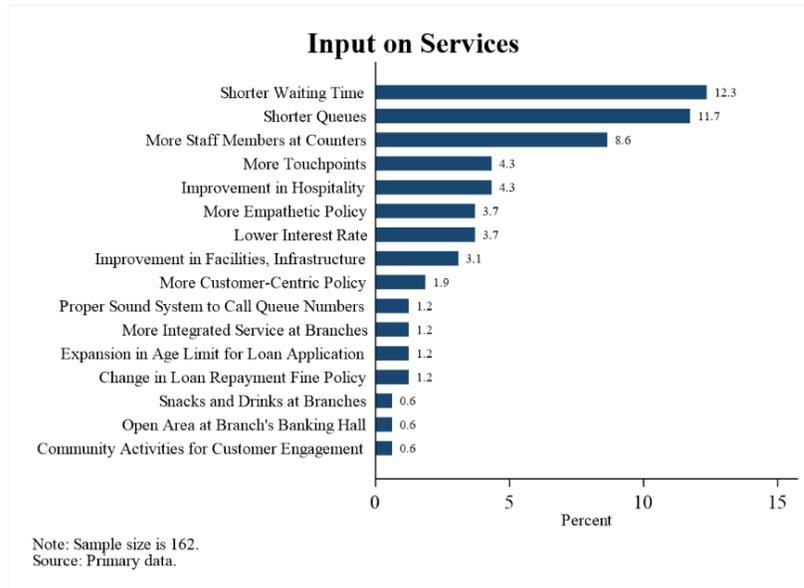

**Figure 8: Input on Services**

From forum group discussions, the bank's reputation was influenced by, among others, customer satisfaction, while customer satisfaction was influenced by several factors, such as transparency of information, the bank's staff's helpfulness in handling customer issues, easiness of interacting with the bank's staff, the bank staff's empathy level in handling customer's issues, and customer's happiness in banking with Bank Jaya Artha. On the other hand, customer satisfaction also affects the customer's likeliness to recommend the bank as a financial services provider and the customer's likeliness to recommend the bank as a lender.

Customers viewed that the bank should be transparent and proactive in delivering information, including but not limited to loan details, such as terms and conditions of loan agreements, the interest rate, the amount of installments customers need to pay every month, the amount of pension salaries that customers receive every month after deductions for installment payments, the maturity of the loans, the number of installments, and the loan repayment fine, which one respondent considered as a risk.

Customers agreed that they were happy in banking with Bank Jaya Artha despite the negative experiences they had encountered, attributing their happiness to the community at the branch office in which they participated. They viewed the community of the bank's pension customers, which is present at every branch and are connected to one another, as an integral part of the bank. They were also happy in banking with Bank Jaya Artha because they could maximize their loan application, and the process to get the loan was faster than that in other banks.

The speedy loan application process and the maximum loan limit together influenced customers' likeliness to recommend the bank as a lender. The bank makes it possible for customers to get loans expeditiously because it requires no medical check-up for the customer's life insurance and no house visit, both as part of the know-your-customer process and it requires no consent from the customer's spouse in the loan application. All three factors also directly influenced customer's likeliness to recommend the bank as a lender.

Customer's likeliness to recommend the bank as a lender was also negatively affected by expensive insurance premiums because of the absence of medical check-ups, high loan repayment fines, the lack of the bank's incentives as a reward for customers bringing in another customer, and by staff members considered unfriendly in providing services to customers. Those negative factors also directly impact customer satisfaction.

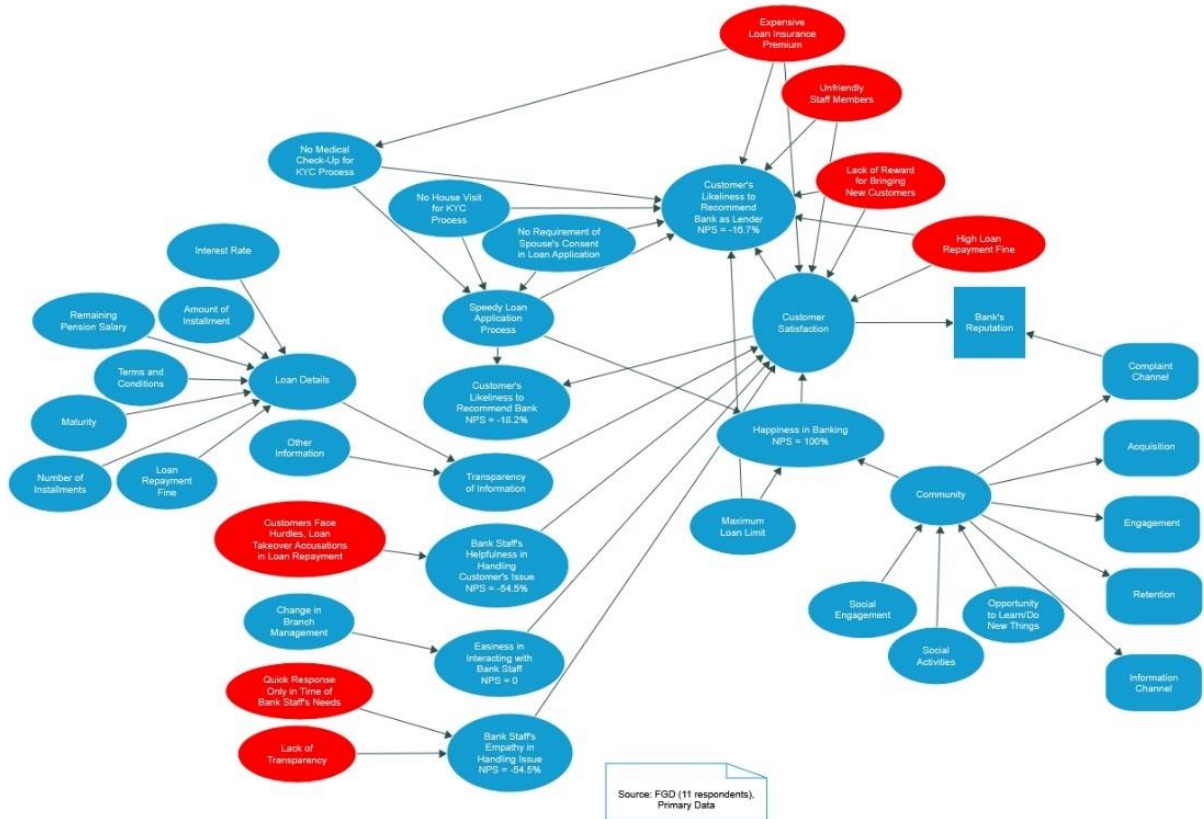

**Figure 9: Bank's Reputation and Affecting Factors**

In the forum group discussions, respondents were also invited to brainstorm possible solutions to the loan repayment fine, and two scenarios were developed.

The first scenario applies to any customers who plan to acquire a new loan. Respondents agreed that the bank must proactively inform customers applying for a new loan about loan terms and conditions, give customers proper time to learn the loan agreement, ensure that they understand what they are about to sign, and secure customers' consent. Meanwhile, in the second scenario, respondents said the bank introduced the new loan repayment fine clause to existing customers, ensured their understanding, and secured their consent before imposition of the new clause. Respondents also agreed that the bank has to refund any existing customers who have been affected by the imposition of the new clause and have paid loan repayment fines.

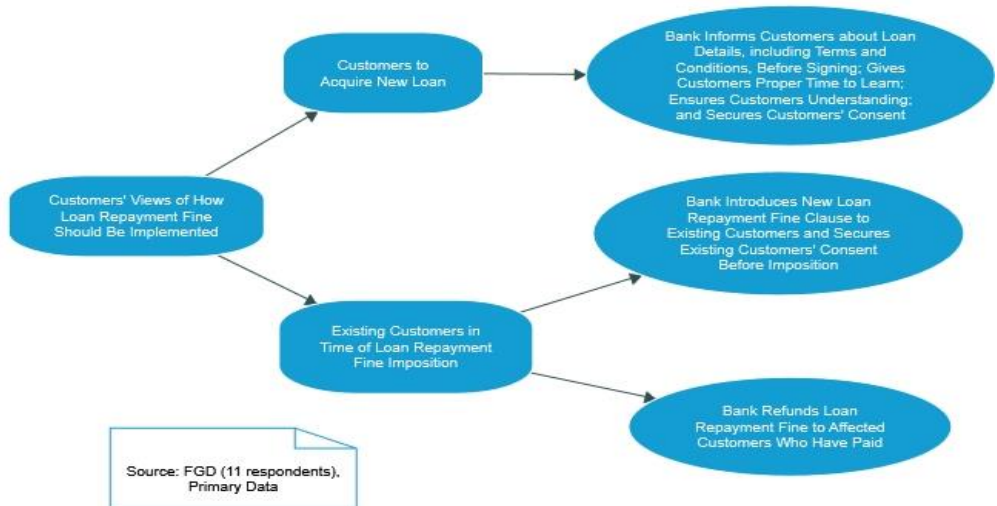

**Figure 10: Proposed Solutions from Ideation with Customers**

## IV. CONCLUSION

There are problems in the way the bank communicates information to customers, particularly information about loan repayment fines that changed in December 2019 and affected existing customers. Problems in communicating loan repayment fines to customers include, but are not limited to, lack of transparency, a reactive approach instead of a proactive one, obscurity of the information, and the time the information is delivered. Transparency of information directly impacts customer satisfaction, which eventually affects the bank's reputation.

A proactive approach, transparency in information delivery, and a more empathetic approach would increase customer satisfaction and eventually benefit the bank.

Customers choose to stay with the bank because of the community, the speedy process to get a loan, and the maximum loan customers can get, despite the negative experiences they encounter. The community serves a pivotal role in anchoring customers because it is also seen as a channel for customers to file a complaint with the bank, as a way to acquire new-to-bank customers, as a platform for customers to engage with each other and for the bank to engage with customers, as means for the bank to retain existing customers, and as a channel for the bank to disseminate information to customers. The community as a complaint channel, however, could directly impact the bank's reputation.

Although customers are generally satisfied with the bank's services, with a 50.6% Net Promoter Score for overall satisfaction, only 26.5% of respondents are likely to recommend Bank Jaya Artha to other people, with a Net Promoter Score for recommendations at -9.3%.

The low recommendation number may stem from customers' experiences with unfriendly staff members, expensive life insurance premiums for loans, lack of reward for bringing in new customers, and high loan repayment fines, as well as from customers' limited social circle and from customers' perception that choice of bank is a personal preference or decision.

In order to become more customer-centric, the bank must proactively inform pension customers about the new loan repayment fine clause, ensure their understanding, obtain their consent, and refund those affected by the new clause that was imposed on customers without their consent based on customers' perspectives.

**Conflict of Interest**
The authors declare that there is no conflict of interest concerning the publishing of this paper.

**Funding Statement**
The authors declare that the research and the publication of this article are self-funded.